\documentstyle[11pt,newpasp]{article}

\begin{document}

\title{The Role of the IILR in the Structure of the Central Gaseous Discs}
\author{C. Yuan and S. Leon}
\affil{Institute of Astronomy \& Astrophysics, Academia Sinica, Nankang,
Taipei, Taiwan 115}

\begin{abstract}
Recent NICMOS observations of the central regions of nearby 
barred galaxies, NGC5383, NGC1530 and NGC1667 by Regan, Sheth and
Mulchaey (1999), all show, in the unsharp mask map, a 
pair of distinct trailing spirals which can be followed 
all the way to the center. This result is in apparent conflict with 
the density wave theory. In this report, we argue that this difficulty 
may be removed if we take the self-gravitation 
of the disc into consideration, or the IILR does not exist in these galaxies.
\end{abstract}


\section{Introduction}

There are two types of rotation curves, one rising rapidly  and the
other relatively slowly from the center. The fast rising rotation curve
usually can host two Lindblad resonances, the outer and the inner (OLR and 
ILR). The slowly rising rotation curve, however, can host one extra inner  
Lindblad resonance. Thus we have an outer inner Lindblad resonance (OILR) 
and an inner inner Lindblad resonance (IILR).
The importance of these Lindblad resonances lies in the fact that
spiral density waves can be excited there by a rotating bar potential. 
These waves dominate the structure and the evolution of the
galactic discs.

Recent NICMOS observations of the central regions of nearby barred 
galaxies of NGC5383, NGC1530, and NGC1667 (Sheth et al 1999; 
Regan et al 1999, Mulchaey and Regan 1997), all show, in the unsharp 
mask map, a pair of distinct trailing spirals in the very center (
within $\sim$500 pc), which can be traced as originating from the
OILR located at a radius of a few kpc from the center.

Preliminary analysis of the rotation curve according to the BIMA
observations (Sheth et al 1999) suggests that NGC5383 has
an IILR. If this is indeed the case, the 
inner spiral structure should be {\em leading} and separated from the 
exterior spiral structure.  The fact that it has a pair of
distinct {\em trailing spiral}, and is connected continuously to the OILR 
spirals, poses a challenge to the theory.  Two explanations are possible
: (1) the IILR is ineffective because its Q-barrier may either be too close 
to the center or not exist, and (2) the rotation curve of NGC5383 
close to the center may not guarantee the existence of an IILR. We shall 
examine both cases here.

\section{The Role of the IILR}

The effectiveness of the IILR in exciting spiral density waves
depends heavily on the location of the Q-barrier, which, in
turn, depends on the self-gravitation of the disk.  The 
self-gravitation will shift the IILR Q-barrier toward the galactic 
center or even into non-existence. In the 
case of NGC5383, a typical disc mass, say, of 200 $M_{\odot}\;pc^{-2}$ 
and typical sound speed of $8\;km/s$, can shift the Q-barrier of
the IILR from 0.5 kpc to 0.2 kpc. If the surface 
density is a little higher and the disc is a little cooler, the 
Q-barrier could be moved "beyond" the center, i.e., non-existence.

Under such a circumstance, the long leading waves excited by the 
bar potential at the IILR may never get reflected or refracted at the 
Q-barrier, Therefore, there would be no reflected short leading spirals 
to be observed. For the same reason, the incoming trailing spirals 
generated at the OILR cannot find the Q-barrier associated with 
the IILR, hence they can go to the center (Yuan and Kuo 1997).

Furthermore, if the Q-barrier were moved to 0.2 kpc, the
disk thickness would be comparable to the the disk radius
(as well as the wavelengths of the density waves). Strong coupling between 
the disc and the thickness would result in strong damping of the 
waves and lead to the effective wave absorption.
Again, it would diminish the Q-barrier's role of reflecting waves, 
hence no short leading spirals.

\section{The Rotation Curves near the Center}

It is well known that the rotation curves near the center of a
disc galaxy cannot be determined accurately, either because of
the large velocity dispersion in the central region or because of 
lack of angular resolution. Yet the Lindblad resonances sensitively
depend on the {\em derivative} of the rotation curve. Although
crude observation data often suggest a rigid body rotation
near the center, thus the existence of an IILR, it is not yet possible
to rule out a Keplerian disk, which has no IILR.
It is entirely possible that NGC5383 does not have an
IILR.  Thus the waves generated at the OILR, can (if viscosity is 
reasonably small) freely march all the way to or close to the center.

\acknowledgments

We are thankful to Drs. M.W. Regan, K. Sheth and J.S. Mulchaey
for letting us use their unpublished observational
results and for their comments on our work.


\begin{references}
\reference Mulchaey, J.S., \& Regan, M.W., 1997, \apj, 482, 135
\reference Regan, W.W., Sheth, K., \& Vogel, S.N., 1999, preprint.
\reference Sheth, K., Regan, M.W., Vogel, S.N., \& Teuben, P.J., 1999,
preprint.
\reference Yuan, C., \& Kuo, C.L., 1997, ApJ, 486, 750.


\end{references}
\end{document}